\documentclass[aps,pre,
twocolumn,
groupedaddress,showpacs]{revtex4-1}

\usepackage{epsfig}
\usepackage{epstopdf}

\usepackage{graphicx}

\usepackage{amsmath}
\usepackage[T1]{fontenc}
\usepackage{float} 
\usepackage{dcolumn}
\usepackage{amsthm}
\usepackage{amssymb}

\usepackage[usenames,dvipsnames]{color}

\definecolor{webgreen}{rgb}{0,0.75,0}
\definecolor{webred}{rgb}{0.75,0,0}
\definecolor{webblue}{rgb}{0,0,0.75}
\definecolor{darkblue}{rgb}{0,0,0.7}
\definecolor{dunkelgrau}{rgb}{0.8,0.8,0.8}
\definecolor{lgray}{rgb}{0.95,0.95,0.95}
\definecolor{lgreen}{rgb}{0.95,1.00,0.90}
\definecolor{lblue}{rgb}{0.9,0.95,1.00}
\definecolor{lred}{rgb}{1.00,0.90,0.80}
\definecolor{shadecolor}{rgb}{1.00,0.92,0.82}

\usepackage{graphicx}

\usepackage{natbib}
\usepackage{multirow}
\usepackage{bbm}

\usepackage{color}

\usepackage[breaklinks]{hyperref}
\hypersetup{colorlinks=true, linkcolor=darkblue, citecolor=darkblue, filecolor=darkblue, urlcolor=darkblue}

\begin{document}


\title{Complexity of Shapiro steps }
\author{ Petar Mali$^{1}$, An\dj ela \v Sakota$^{1}$, Jasmina Teki\' c$^2$, Slobodan Rado\v sevi\' c$^{1}$, Milan Panti\' c$^{1}$, Milica Pavkov - Hrvojevi\' c$^{1}$}
\affiliation{$^1$ Department of Physics, Faculty of Science, University of Novi Sad,
Trg Dositeja Obradovi\' ca 4, 21000 Novi Sad, Serbia}
\affiliation{$^2$"Vin\v ca" Institute of Nuclear Sciences,
Laboratory for Theoretical and Condensed Matter Physics - 020,
University of Belgrade, PO Box 522, 11001 Belgrade, Serbia}

\date{\today}


\begin{abstract}
We demonstrate on the example of the dc+ac driven overdamped Frenkel-Kontorova model that an easily calculable measure of complexity can be used for the examination of Shapiro steps in presence of thermal noise.
In real systems, thermal noise causes melting or even disappearance of Shapiro steps, which makes their analysis 
in the standard way from the response function difficult.
Unlike in the conventional approach, here,
by calculating the Kolmogorov complexity of certain areas in the response function we were able to detect Shapiro steps, measure their size with desired precision and examine their temperature dependence.
The aim of this work is to provide scientists, particularly experimentalists, an unconventional but a practical and easy tool for examination of Shapiro steps in real systems.
\end{abstract}
\pacs{05.45.-a; 45.05.+x; 71.45.Lr; 74.81.Fa} \maketitle

\section{Introduction}
\label{intro}

It is well known that dynamical systems with competing frequencies 
can exhibit some form of frequency locking and Shapiro steps.
Since their discovery in Josephson junctions, Shapiro steps have been widely studied phenomenon in all kinds of nonlinear systems  
from charge density waves \cite{Grun, GrunPR, Thorne, Hund, Sinch}, and
Josephson junctions~\cite{Dub, Sel, Lim, Shukrinov, Shukrinov2} to colloidal systems~\cite{Nature, NJP} and superconducting nanowires \cite{Dins, Bae, Rid}.
However, in experiments these systems are exposed to various environmental effects among which certainly the most significant is the thermal noise.
Thermal noise  greatly affects mode locking by melting and changing properties of Shapiro steps such as their amplitude \cite{ACT, ACTr} and frequency dependence \cite{ACTr, ACTn}. 
It is, therefore, very difficult or even impossible to detect, measure or control them, and
often, different methods have to be developed in order to address this problem. 
In charge density wave systems and Josephson junctions, for example, instead of current-voltage  characteristics, where steps are hardly visible due to noise, differential resistance is used for their analysis~\cite {Hund, Sinch, Dub}.
On the other hand, Shapiro steps are ideally suited for realizing a voltage standard in different devices that are used in various
technological applications \cite{Buckel}. 
Therefore,  an understanding  of Shapiro steps as a phenomenon and control of their
behavior in real systems is of great importance.  

In the meanwhile, in the theory of nonlinear dynamical systems an interesting and easily applicable tool known as complexity measure or Kolmogorov complexity (KC)~\cite{Kolmogorov, Ziv} has been proposed.
Today it is closely related to the information theory~\cite{Kolmogorov, Ziv, Kaspar, Kov}, and applied in many other fields such as 
hydrology~\cite{Mih}, climatology~\cite {Mih2,Gordan}, etc.
The complexity measure represents a useful quantity, which characterizes spatiotemporal patterns and provides fine measure of order, i.e., periodic motion and chaotic behavior. 
Nonlinear systems, which exhibit Shapiro steps, 
besides the quasiperiodic and periodic motion can also exhibit a transition to chaos.
It is, therefore, naturally to ask whether a complexity measure could be a convenient tool for the studies of Shapiro steps.

In this paper we will apply  on the studies of Shapiro steps a method very different from the conventional ones, the Kolmogorov complexity. 
In particular, we will consider the dc+ac driven overdamped Frenkel-Kontorova (FK) model with deformable substrate potential in the presence of noise. 
The standard FK model represents a chain of harmonically interacting identical particles subjected to sinusoidal substrate potential \cite{OBBook,ACFK}. 
It can describe various commensurate and incommensurate structures \cite{Obri1,Obri2}, which exhibit very rich dynamics under external forces.
When external dc and ac forces are applied, the mode locking appears due to resonance between the frequency of particle motion over the substrate potential and the frequency of the external ac force \cite{ACFK, Flor, Falo, FlorAP}.
This effect is characterized as a staircase of resonant steps, i.e., Shapiro steps in 
the plot of average velocity as a function of average driving force $\bar{v}(\bar{F})$.
The steps are called harmonic if the locking appears at integer values of ac frequencies or subharmonic if it appears at noninteger rational ones.
Although it was successful 
in describing harmonic steps, the standard ac+dc driven overdamped FK model can not be used for modeling phenomena related to subharmonic steps \cite{ACFK,ACDS}.
Namely, subharmonic steps do not exist in  commensurate structures with integer values of winding numbers while for  noninteger values their size is too small, 
which makes their analysis very difficult \cite{Falo,Rene,Wu}. To overcome this,
some generalizations of standard FK model are necessary.
For instance, large subharmonic steps  in staircase like response
appear in FK model with asymmetric deformable potential \cite{Bambi,ACr}.  The fact that subharmonic
steps are present even in the case of integer value of winding number $\omega = 1$ \cite{Sat} indicates that, by choosing
this type of potential, additional effective degrees of freedom are induced in the system.
Therefore, generalization of the FK model by using some form of deformable substrate potential  
provides a good framework for studies of subharmonic mode locking~\cite{ACFK, ACr, ACP}.

If noise is present in the system, at certain temperature, the response function $\bar{v}(\bar{F})$ will be substantially affected. 
All steps will start melting, and while
some will be more robust, the others might disappear completely \cite{ACFK, ACT, ACTr}. Therefore,  
it is often hard to get any information about resonances just from the observation of response function $\bar{v}(\bar{F})$.
We will further introduce a method, which can overcome that difficulty, and by using KC, examine in detail Shapiro steps in the presence of noise.

\section{Model and method}
\label{model}

We consider the dynamics of coupled harmonic oscillators $u_l$, subjected to asymmetric deformable substrate potential~\cite{Peyrard}:
\begin{equation}
\label{V}
V(u)=\frac{K}{4 \pi^2}\frac{(1-r^2)^2 \big[1-\cos(2\pi u) \big]}{\big[1+r^2+2r\cos(\pi u) \big]^2},
\end{equation}
where $K$ is the pinning strength and $r$  is deformation parameter ($-1< r <1)$.
By changing  $r$, the potential can be tuned in a very fine way,
from the simple sinusoidal one for $r=0$  and to a deformable one for $0<|r|<1$.
The total potential energy of such system is 
\begin{equation}
H = \sum_{l} \left( V(u_l) + W (u_{l+1}-u_l)  \right),
\end{equation}
where 
\begin{equation}
W (u_{l+1}-u_l) = \frac 12 \left( u_{l+1}-u_l  \right)^2
\end{equation}
represents harmonic coupling between  neighboring particles \cite{Obri1,Obri2}.
The system is driven by dc and ac  forces,
%
$F(t)=F_{\mathsf{dc}} + F_{\mathsf{ac}}\cos(2\pi\nu_0t)$,
%
where $F_{\mathsf{ac}}$ and $\nu_0$ are amplitude and frequency of ac force 
which, in the overdamped limit,  leads to the system of equations of motion:
\begin{equation}
\label{u}
\dot{u}_l=u_{l+1}+u_{l-1}-2u_l-\frac{\partial V}{\partial u_l}+F_{\mathsf{dc}}+F_{\mathsf{ac}}\cos (2\pi \nu_0 t)+L_l(t), 
\end{equation}
where $l=1,...,N$, and $N$ is the number of particles, and  $u_{N+1}=u_1$. 
The noise term is chosen as Gaussian white noise which satisfies
$\langle L_l(t)L_{l'}(t') \rangle=2T\delta_{l,l'}\delta(t-t')$.

When the system is driven by a periodic force, 
the competition between the frequency $\nu _0$ of the external periodic (ac) force and
the characteristic frequency of the particle motion over the periodic substrate potential driven by
the average force $\bar F=F_{\mathsf{dc}}$
results in the appearance of dynamical mode locking.
The solution of the system (\ref{u}) is called resonant if average velocity $\bar{v}$ satisfies the relation \cite{ACDS}:
\begin{equation}
\label{v}
\bar{v}=\left(i \pm \frac{1}{m\pm \frac{1}{n\pm \frac {1}{p\pm ...}}} \right) \omega\nu_0,
\end{equation}
where $i, m, n, p$ are integers.
The first level terms, which involve only $i$, represent harmonic steps, whereas the other terms describe subharmonic steps. 
The system of equations (\ref{u}) has been numerically integrated using periodic boundary conditions for
the commensurate structure $\omega =\frac 12 $  with two particles per potential well.
The time step used in the simulations
was $0.001 \tau$, and a time interval of $100 \tau$	was used as a
relaxation time to allow the system to reach the steady state~\cite{ACT, ACTr}.
The dc force was varied with steps $10^{-5}$ and $10^{-6}$.

Before we focus on the Shapiro steps, let us shortly introduce the Kolmogorov complexity.
In general, the KC of a finite object is defined as the length of the shortest effective binary description of that object.
In other words, if we have some binary sequence of the length $n$, then the measure of complexity, $c(n)$, of that sequence is the number of
patterns or parts required to reproduce that sequence.
Calculation of $c(n)$ is performed by Lempel--Ziv algorithm (detailed explanation can be found in \cite{Ziv, Kaspar, Mih}).
We shall describe the algorithm first and then illustrate it with a specific
example. 

If we have some sequence $(x_1,..., x_n)$ of zeros and ones, the algorithm which determines the patterns, which form this sequence, consists of the following steps:
(1) for a given sequence of zeros and ones, the first digit is always the first pattern; 
(2) define the sequence $S=x_1,..., x_k\cdot $, which contains the first pattern and grows until the whole sequence is analyzed; 
(3) in order to check whether the rest of sequence $(x_{k+1},..., x_n)$
can be reconstructed by simple copying (or whether one has to insert new digits), define sequence $Q\equiv x_{k+1}$ by adding a new digit; 
(4) define sequence $SQ$ by adding $S$ and $Q$; 
(5) form the sequence $SQ\pi $ by removing the last digit of sequence
$SQ$ and examine whether $Q$ is part of the vocabulary $V(SQ\pi )$, if it is not, then sequence $Q$ is the new pattern, and this new pattern is added to the list of known patterns called vocabulary $R$.
Sequence $SQ$ becomes new sequence $S$, while $Q$ is emptied and ready for further testing.
If on the other hand
$Q$ is part of $SQ\pi $, repeat the process until the mentioned condition is satisfied.

For an illustration let us consider the binary sequence $11010001$.
The first digit is the first pattern $R=1\cdot $. 
Since the first two digit are both $1$, we have that $S=1$, $Q=1$, which leads to $SQ=11$ and $SQ\pi =1$. 
In the following step we ask whether $Q$ is contained in the vocabulary $V(SQ\pi)$. 
Since $Q=1$ is the part of $SQ\pi =1$, we add the third digit to $Q$ so that $Q=10$, while $S=1$ and repeat the procedure.
The algorithm can be written as:
\begin{itemize}
\item \noindent
$S=1, Q=1, SQ=11, SQ\pi=1, \\
Q \in V(SQ\pi) \rightarrow  R=1 \cdot 1$
\item 
$S=1, Q=10, SQ=110, SQ\pi=11, \\
Q \notin V(SQ\pi) \rightarrow R=1 \cdot 10 \ \cdot $
\item 
$S=110, Q=1, SQ=1101, SQ\pi=110, \\
Q\in V(SQ\pi) \rightarrow R=1 \cdot 10 \cdot 1$ 
\item 
$S=110, Q=10, SQ=11010, SQ\pi=1101, \\
Q\in V(SQ\pi) \rightarrow R=1 \cdot 10 \cdot 10$
\item 
$S=110, Q=100, SQ=110100, SQ\pi=11010, \\
Q\notin V(SQ\pi) \rightarrow R=1 \cdot 10 \cdot 100 \ \cdot$  
\item 
$S=110100, Q=0, SQ=1101000, SQ\pi=110100,
Q \in V(SQ\pi) \rightarrow R=1 \cdot 10 \cdot 100 \cdot 0$
\item 
$S=110100, Q=01, SQ=11010001, SQ\pi=1101000,
Q \in V(SQ\pi) \rightarrow R=1 \cdot 10 \cdot 100 \cdot 01$  
\end{itemize}
According to Lempel--Ziv algorithm the above sequence can be written as $1 \cdot 10 \cdot 100 \cdot 01$.
The measure of complexity $c(n)$ is simply the number of parts separated by dots, i.e., the pattern counter, 
which in this case is $c=4$.

The application of KC on the analysis of Shapiro steps 
in the presence of thermal noise,
in principle, requires comparison of  sequences which
differ in length. This is due to the fact that  Shapiro steps
do not have equal widths.
Previous remark holds for Shapiro steps
observed in experiments or  in numerical
simulations.
In our analysis of Shapiro steps we will further use the normalized measure of complexity given as \cite{Ziv, Kaspar}:
\begin{equation} 
\label{CK}
C_{\mathsf{K}}=\frac{c(n)}{b(n)},  
\end{equation}
where $b(n) = \frac{n}{\mbox{log}_2n}$ represents the asymptotic value of $c(n)$ \cite{Kaspar}.
This means that for sequence of infinite length ($n \to \infty$),  Kolomogorov complexity (\ref{CK})
takes values  $C_{\mathsf{K}} \in [0,1]$, i.e.
$0 \leq \lim_{n \to \infty} [c(n)/b(n)] \leq 1$.
However, for sequences of finite length,  the values of KC
can exceed 1\footnote{\label{FN} According to (\ref{CK}), the value of KC for the 
binary sequence $11010001$ would therefore be $C_{\mathsf{K}}=1.5$. 
However, the given sequence is too short, $n=8$, and consequently, the obtained result for $C_{\mathsf{K}}$ is not relevant.}.
Since it was shown in \cite{Kaspar} that the limit  
$\lim_{n \to \infty} [c(n)/b(n)] $  is reached within
$5\%$ for $n > 10^3$,
all calculations of the Shapiro steps complexity in next section were performed on sequences of size $n>1000$.
In numerical simulations this length can be easily 
reached with suitable choice of force step $\delta F_{\mathsf{dc}}$.

\section{Results} \label{Res}

\begin{figure}[ht] 
\includegraphics[width=8 cm]{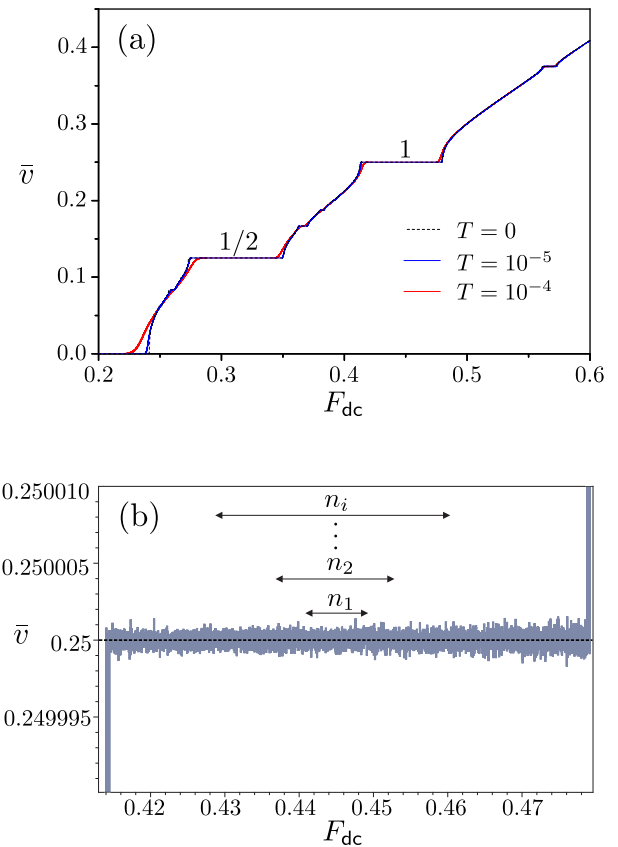}
\centering
\caption{\label{Fig1}
(Color online) (a) Average velocity $\bar{v}$ as a function of the average driving force $F_{\mathsf{dc}}$ for $K=4$, $\omega=1/2$, $r=0.2$, $\nu_0=0.5$, and tree different values of temperature
 $T=0$, $T=10^{-4}$, and $T=10^{-5}$.
(b) High resolution plot of the first harmonic step in (a). 
$F_{\mathsf{dc}}$  was varied with step $\delta F_{\mathsf{dc}} = 10^{-5}$ and $n_1$, $n_2$,... $n_i$ denote lengths of sequences for which $C_{\mathsf{K}}$ was calculated. 
Dashed black line at $\bar{v}=0.25$ represents the first harmonic step at $T=0$ and  the blue line represents
the first harmonic step at $T=10^{-5}$. }
\end{figure}

Influence of temperature on the response function $\bar{v}(F_{\mathsf{dc}})$ is presented in Fig. \ref{Fig1}.
The presence of thermal noise causes melting of Shapiro steps, which become more and more rounded as the temperature increases as can be seen  in Fig. \ref{Fig1}(a); 
meanwhile the critical depinning force decreases~\cite{ACFK, ACT, ACTr}. 
At very high temperature the pinning potential has no influence and the system becomes the system of free particles~\cite{ACFK, ACT}.
At $T=0$, the step can be analyzed easily, however,
if $T\neq 0$, just a simple task such as measuring the step size can not be done accurately enough since it is impossible to determine exactly at which value of $F_{\mathsf{dc}}$ the step begins or ends.
In conventional approach, some criterion needs to be always introduced in order to determine the beginning and the end of each step.
In our previous works we considered the system to be on the step if the changes of $\bar v(\bar F)$ are
less than $0.1\% $~\cite{ACT, ACTr, ACTn}.
When steps are large and the temperature is low this does not represent a difficulty.  
However, for smaller subharmonic steps or at the higher temperature the error could be the order of step or the order of the fluctuations of resonant value $\bar v (T\neq 0)$ around $\bar v (T=0)$, which makes analysis of subharmonic steps challenging. 

\begin{figure}[ht] 
\includegraphics[width=8 cm]{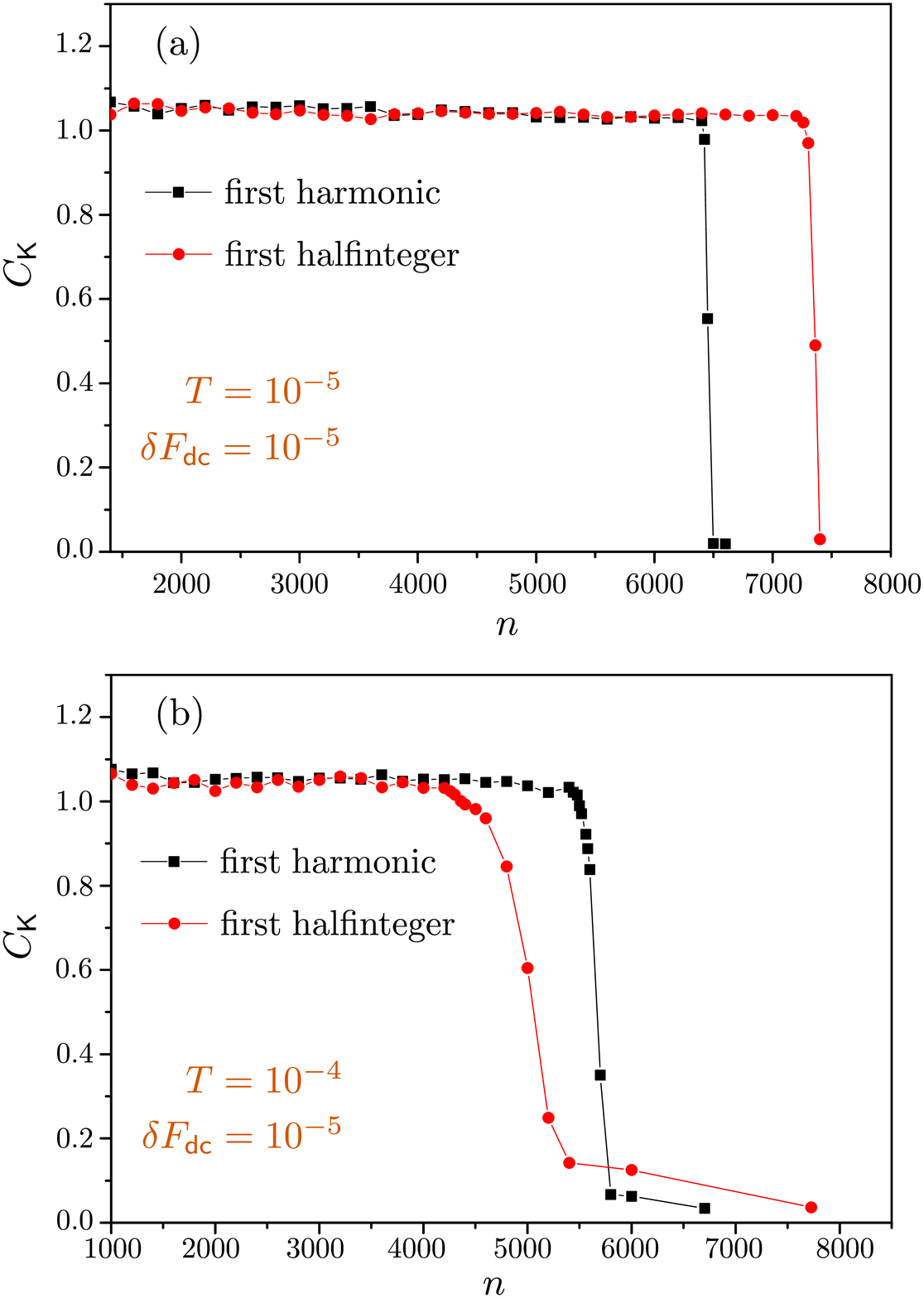}
\centering
\caption{\label{Fig2}
(Color online) Kolmogorov complexity $C_{\mathsf{K}}$ of the first harmonic and halfinteger step as a function of the length of their corresponding sequences for $K=4$, $\omega=1/2$,  $r=0.2$, $F_{\mathsf{ac}}=0.2$, $\nu_0=0.5$ and $T=10^{-5}$, and $10^{-4}$ in (a) and (b) respectively. 
The dc force was varied with step $\delta F_{\mathsf{dc}}=10^{-5}$.}
\end{figure}

From Fig. \ref{Fig1}(b) at $T\neq 0$
it is evident that
the function $\bar{v}(F_{\mathsf{dc}})$ oscillates around the value $\bar{v}=0.25$, which corresponds to the position of first harmonic step at $T=0$. 
In order to calculate $C_{\mathsf{K}}$, we need to transform the values for average velocity into a corresponding binary string. 
To achieve this we define:
\begin{equation}
\label{vN}
\langle \bar{v} \rangle=\frac{1}{n}\sum^n_{i=1}\bar{v}_i, 
\end{equation}
where $\bar{v}_i$ corresponds to values of average force taken with step $\delta F_{\mathsf{dc}}$. 
Then we prescribe values $0$ or $1$ to particular $\bar{v}_i$ according to:
\begin{equation}
\label{vi} 
\bar{v}_i \rightarrow \left\{\begin{array}{rl}
1,  & \bar{v}_i \geq \langle \bar{v} \rangle\\
0, & \bar{v}_i < \langle \bar{v} \rangle.
\end{array}
\right. 
\end{equation}
Using binary sequences obtained in this way, starting from the mid point of step at $T=0$ in Fig. \ref{Fig1}(b), we calculate the KC for sequences with increasing number of terms $n$. 
The output from these calculations are shown in Fig. \ref{Fig2} for two different values of temperature.
As we can see in Fig. \ref{Fig2}(a), the existence of  Shapiro step is obvious, i.e., the Kolmogorov complexity $C_{\mathsf{K}}$ determines the range of parameter $n$ for which $C_{\mathsf{K}}\geq 1$. 
Therefore, we define a criterion that when $C_{\mathsf{K}}$ becomes less than $1$, the system is no longer on the step, i.e., locked.
Following this, we can determine very accurately the step width.
If dc force was varied with some step $\delta F_{\mathsf{dc}}$, then the steps width can be determined as
\begin{equation} 
\label{DF}
\Delta F=n_{\mathsf{max}}\, \delta F_{\mathsf{dc}},  
\end{equation}
where $n_{\mathsf{max}}$ is the maximal value of $n$ for which $C_{\mathsf{K}} \geq 1$. 
The set of parameters in Fig. \ref{Fig2}
was chosen in such manner that the first halfinteger is larger than the first harmonic step at $T=0$.
If the temperature is increased, the steps are melting further as can be seen in Fig. \ref{Fig2}(b). 
Here, $C_{K}$ clearly shows not only the size of steps but it also reveals their robustness: harmonic step is more robust under the influence of noise, which is in agreement with \cite{ACTr}.

\begin{figure}[ht] 
\includegraphics[width=8 cm]{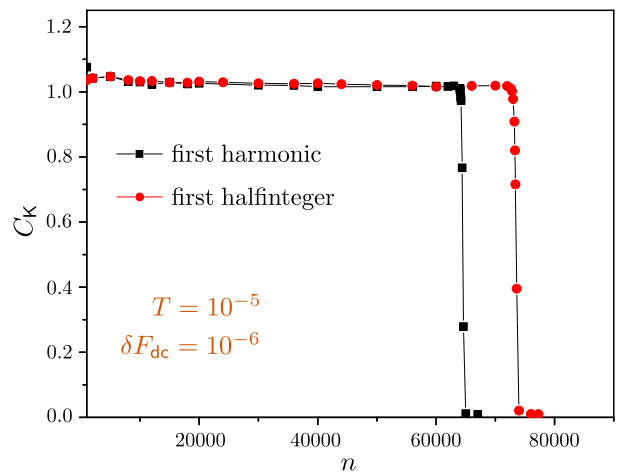}
\centering
\caption{\label{Fig3}
(Color online) Kolmogorov complexity $C_{\mathsf{K}}$ of the first harmonic and halfinteger step as a function of the length of their corresponding sequences for $\delta F_{\mathsf{dc}}=10^{-6}$.
The rest of parameters is the same as in Fig. \ref{Fig2}.
}
\end{figure}

If we increase $n$ ten times by decreasing the force step to $\delta F_{\mathsf{dc}}=10^{-6}$, we can see in Fig. \ref {Fig3} that the results remain unchanged compared to the results in Fig. \ref{Fig2}(a). 
The length of sequence $n$ is essential for the precision of $C_{\mathsf{K}}$, and it is determined by the resolution ($\delta F_{\mathsf{dc}}$) in which plots for $\bar v(F_{\mathsf{dc}})$ are produced.  
Thus, depending on the size of steps and how precisely we want to measure them, the resolution in which steps are obtained should be adjusted accordingly.
In our case, we can see that the resolution, which corresponds to $\delta F_{\mathsf{dc}}=10^{-5}$, provides accurate results for large harmonic or halfintiger steps.
Detection and measurement of higher order subharmonic steps, which are typically very small, sometimes requires higher resolution, i.e. larger $n$. \textit{However, we checked that for the force step $\delta F_{\mathsf{dc}}=10^{-3}$, Shapiro step widths agree within $1\%$ with those obtained with high resolution $\delta F_{\mathsf{dc}}=10^{-5}$ (see Appendix). 
This is due to the fact that decreasing  $n$  induce variations of $C_{\mathsf{K}}$ in such manner
that complexity is larger for shorter
sequences when the system is on the step. Since proposed measure works
even for shorter sequences, it could be of relevance in experimental studies.}
In our calculation of $C_{\mathsf{K}}$, we started always from the midpoint of the steps as it was shown in Fig. \ref{Fig1}(b) since they melt symmetrically on the both sides, but in general, any point on the step can be chosen as the starting point, and by the expansion of $n$, the edges of each step can be precisely determined.

\begin{figure}[ht] 
\includegraphics[width=8 cm]{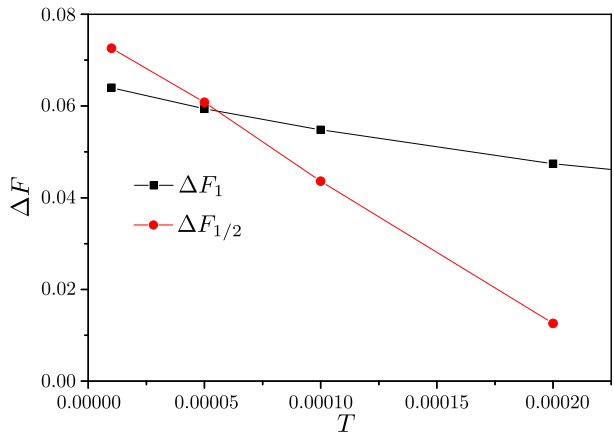}
\centering
\caption{\label{Fig4}
(Color online) The width of first harmonic $\Delta F_1$ and halfinteger step $\Delta F_{1/2}$ calculated from Kolmogorov complexity for $K=4$, $\omega=1/2$, $r=0.2$, $F_{\mathsf{ac}}=0.2$, $\nu_0=0.5$ and $T=10^{-5}$. 
The dc force was varied with step $\delta F_{\mathsf{dc}}=10^{-5}$. }
\end{figure}

Finally, we use KC to examine the temperature  dependence of Shapiro steps.
In Fig.\ \ref{Fig4}, the temperature dependence of the first harmonic and halfinteger step is presented.
The results clearly show that though the halfinteger step melts much faster, there is still a region of temperature for which halfinteger step is larger than harmonic one, 
which is again in very good agreement with the results of temperature dependence obtain in conventional way \cite{ACTr}.

\section{Conclusion}
\label{concl}

In this work we applied Kolmogorov complexity on the studies of Shapiro steps in the presence of thermal noise.
Conventional analysis of steps from the
response function requires some consistent
criterion of measure. However, this leads to larger
errors in the case of smaller steps or
higher temperatures. In contrast, using KC, we
were  able to detect and measure steps very
precisely regardless of the system parameters.
Though, we presented results for only harmonic and halfinteger steps, this technique could be also easily applied on the studies of even smallest subharmonic steps. 
It is important to note that this technique is also  model independent, and
since it relies only on the output data it could be equally well used for experimental and theoretical investigation of Shapiro steps in any system. 
For instance, the critical current  in irradiated Josephson junctions can be defined as the applied current at which a
finite voltage response first appears, and corresponds to a
critical depinning force $F_{\mathsf{c}}$, whereas current step $\delta I_{\mathsf{dc}}$
is analogous to the force step $\delta F_{\mathsf{dc}}$  \cite{Reichard}. Also,
widths of Shapiro steps $\Delta V$ in irradiated Josephson junction systems
correspond to $\Delta F$ considered in this paper \cite{ACDS,EPL}.

Shapiro steps 
have wide applications
and Kolmogorov complexity could be a practical tool 
which can improve measurments of their size in presence
of thermal noise.
To illustrate that, let us look at the recent studies that attracted great attention: application of Josephson junctions in detection of Majorana fermions~\cite{Jiang, Houz, ShukrinovPRB2}.
In Ref. \cite{ShukrinovPRB2}, it was shown that 
the current-voltage characteristics exhibits  odd Shapiro steps, where particular sequence of subharmonic steps in the devil's staircase structure represents a signature of the Majorana states. 
Thus, detection of Majorana fermions requires very precise observation and measurement of subharmonic Shapiro steps, which in experiments could be a real problem due to noise
(all analysis in Ref. \cite{ShukrinovPRB2} were performed at the zero temperature).
We believe that application of KC could help in the solution of this and similar problems.
We have been focused here only on Shapiro steps in the presence of thermal effects, but KC could be used for studies of other effects among which certainly interesting is the chaos \cite{Shukrinov, Shukrinov, ACm}.
Application of Kolmogorov complexity on the studies of chaotic behavior in systems which exhibit Shapiro steps are part of our future studies and will be published separately. 
\vspace*{0.3cm}

\begin{acknowledgments}
We wish to express our gratitude to Jovan Odavi\' c and Gordan Mimi\' c for helpful discussions.
This work was supported by the Serbian Ministry of
Education and Science under Contracts No. OI-171009 and No. III-45010 and by the Provincial 
Secretariat for High Education and Scientific Research of Vojvodina (Project No. APV 114-451-2201).
\end{acknowledgments}

\bibliography{Paper_Mali}

\vspace*{1cm}
\appendix*

\section{Complexity of shorter sequences }

\begin{figure}[ht] 
\includegraphics[width=8 cm]{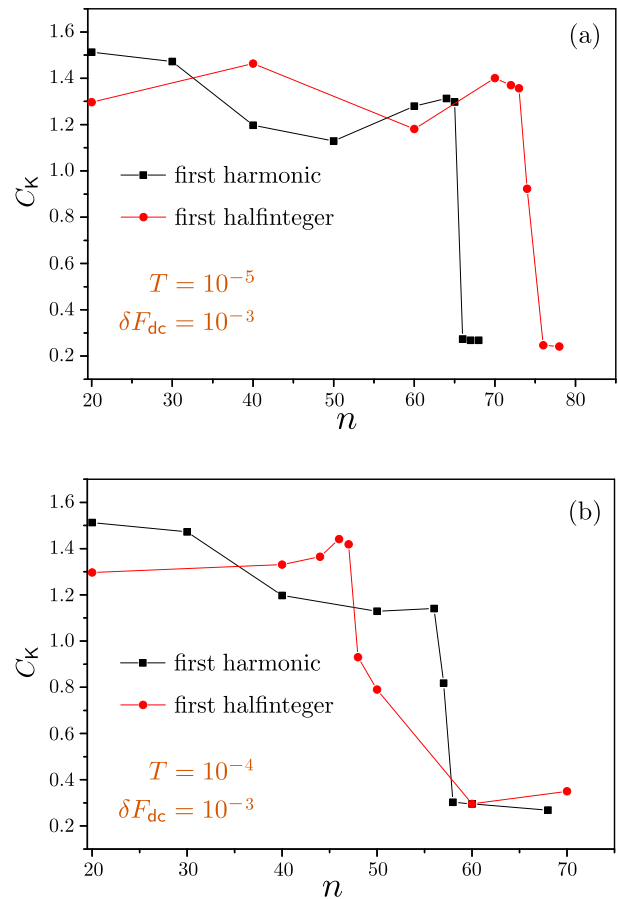}
\centering
\caption{\label{Fig5}
(Color online) Kolmogorov complexity $C_{\mathsf{K}}$ of first harmonic  and halfinteger step  calculated for  $K=4$, 
$\omega=1/2$, $r=0.2$, $F_{\mathsf{ac}}=0.2$, $\nu_0=0.5$ and two temperatures a) $T=10^{-5}$ and b) $T=10^{-4}$. 
The dc force was varied with step $\delta F_{\mathsf{dc}}=10^{-3}$. }
\end{figure}

For the force step $\delta F_{\mathsf{dc}} = 10^{-3}$, we 
obtain Shapiro step widths within $1\%$ of values calculated
with $\delta F_{\mathsf{dc}} = 10^{-5}$  by using relation
$\Delta F = n_{\mathsf{max}} \delta F_{\mathsf{dc}}$ (see Fig.\ \ref{Fig5}).


\end{document}